\documentclass[twocolumn,amsmath,amssymb,aps]{revtex4-1}
\usepackage{graphicx}

\begin{document}

\title{Machine learning design of a trapped-ion quantum spin simulator}

\author{Yi Hong Teoh$^1$, Marina Drygala$^{1,2}$, Roger G. Melko$^{1,2}$, Rajibul Islam$^1$}
\address{$^1$Institute of Quantum Computing and Department of Physics and Astronomy, University of Waterloo, 200 University Ave. West, Waterloo, Ontario N2L 3G1, Canada}
\address{$^2$Perimeter Institute for Theoretical Physics, Waterloo, Ontario N2L 2Y5, Canada}


\begin{abstract}
Trapped ions have emerged as one of the highest quality platforms for the quantum simulation of 
interacting spin models of interest to various fields of physics.  In such simulators, two effective spins 
can be made to interact with arbitrary strengths by coupling to the collective vibrational or phonon states of ions, controlled by precisely tuned laser beams.  However, the task of determining laser control parameters required for a given 
spin-spin interaction graph is a type of {\it inverse problem}, which can be highly mathematically complex.
In this paper, we adapt a modern machine learning technique developed for similar inverse problems to 
the task of finding the laser control parameters for a number of interaction graphs.
We demonstrate that typical graphs, forming regular lattices of interest to physicists, can easily be produced for up to 50 ions using a single GPU workstation. The scaling of the machine learning method suggests that this can be expanded to hundreds of ions
with moderate additional computational effort.
\end{abstract}

\maketitle

\section{Introduction}
Quantum simulators are experimentally well-controlled devices that can be used to emulate quantum many-body systems that may otherwise be intractable by classical computation \cite{Feynman1982simulating,Cirac2012goals,Georgescu2014quantum}. Among several hardware platforms \cite{Cirac2012goals,Aspuru-Guzik2012photonic,Bloch2012quantum,Blatt2012quantum} for quantum simulation, trapped ions have proven to be extremely versatile, capable in particular of preparing many-body spin systems of interest to quantum information and condensed matter physicists \cite{Friedenauer2008simulating,Kim2010quantum,Barreiro2011open-system,Britton2012engineered}. Spin-states, encoded in optical or hyperfine states of individual ions, can interact via phonon modes \cite{Cirac1995quantum, Sorensen1999quantum} arising from the interplay of the trapping potential and long-range Coulomb interactions. The interactions are inherently long-ranged, and the spin-spin interaction graph can be engineered arbitrarily \cite{Korenblit2012quantum}, in principle, by controlling spin-phonon couplings that are (virtually) excited by precisely tuned laser beams. The ability to engineer an arbitrary spin-spin interaction graph opens up exciting possibilities for quantum simulation - such as the simulation of high energy physics problems \cite{Muschik2017lattice}, or the analog quantum simulation of interacting spins on a re-programmable lattice geometry in arbitrary dimensions. Analog simulators, where the target Hamiltonian is engineered in continuous time, are resilient to digital or Trotterization errors \cite{Blatt2012quantum} encountered in a digital quantum simulation, where the target unitary evolution is engineered using discrete quantum logic gates. An engineered interaction graph may also be used as a starting point for analog-digital hybrid quantum simulation protocols \cite{Rajabi2019dynamical,Hayes2014programmable} that offer advantages over both analog and digital quantum simulations. The capability to simulate two and higher dimensional lattice geometries greatly extends the usefulness of conventional trapped ion systems, where $N>100$ ions have been trapped in a linear geometry \cite{Pagano2018cryogenic}. 

However, with the ability to create arbitrary interaction graphs comes the inherent difficulty in tuning the many experimental parameters (such as laser frequencies and intensities) in the quantum simulator.  In fact, the task of determining control parameters may be analytically ill-posed, requiring numerical regularizing algorithms to find approximate pseudo-solutions. In a previously proposed approach \cite{Korenblit2012quantum}, individual nonlinear optimizations of the $N^2$ control parameters were implemented for a static interaction graph. However, a general framework to efficiently extract the experimental control parameters for an arbitrary interaction graph would be desirable for simulating problems with dynamical interactions, such as quench and transport problems. In this paper, we demonstrate how modern, powerful machine learning techniques based on artificial neural networks can be used to find practical, verifiable, and useful solutions to this problem. A properly trained neural network can, in approximately 1 ms on conventional computer hardware, produce experimental control parameters to achieve a range of target interaction graphs. We illustrate the technique by finding the laser control parameters that produce highly accurate spin-spin interaction graphs in a linear chain of ions.  A number of cases of physical interest for quantum spin simulation are produced, including square, triangular, kagome and cubic lattices. With modern experiments employing individual control of trapped-ion qubits and phonon modes that connect them, our protocol is directly relevant to the rapid progress of the field of practical quantum information processing.

\section{A neural network for the inverse problem}
\begin{figure}[ht]
    \centering
    \includegraphics[width=\columnwidth]{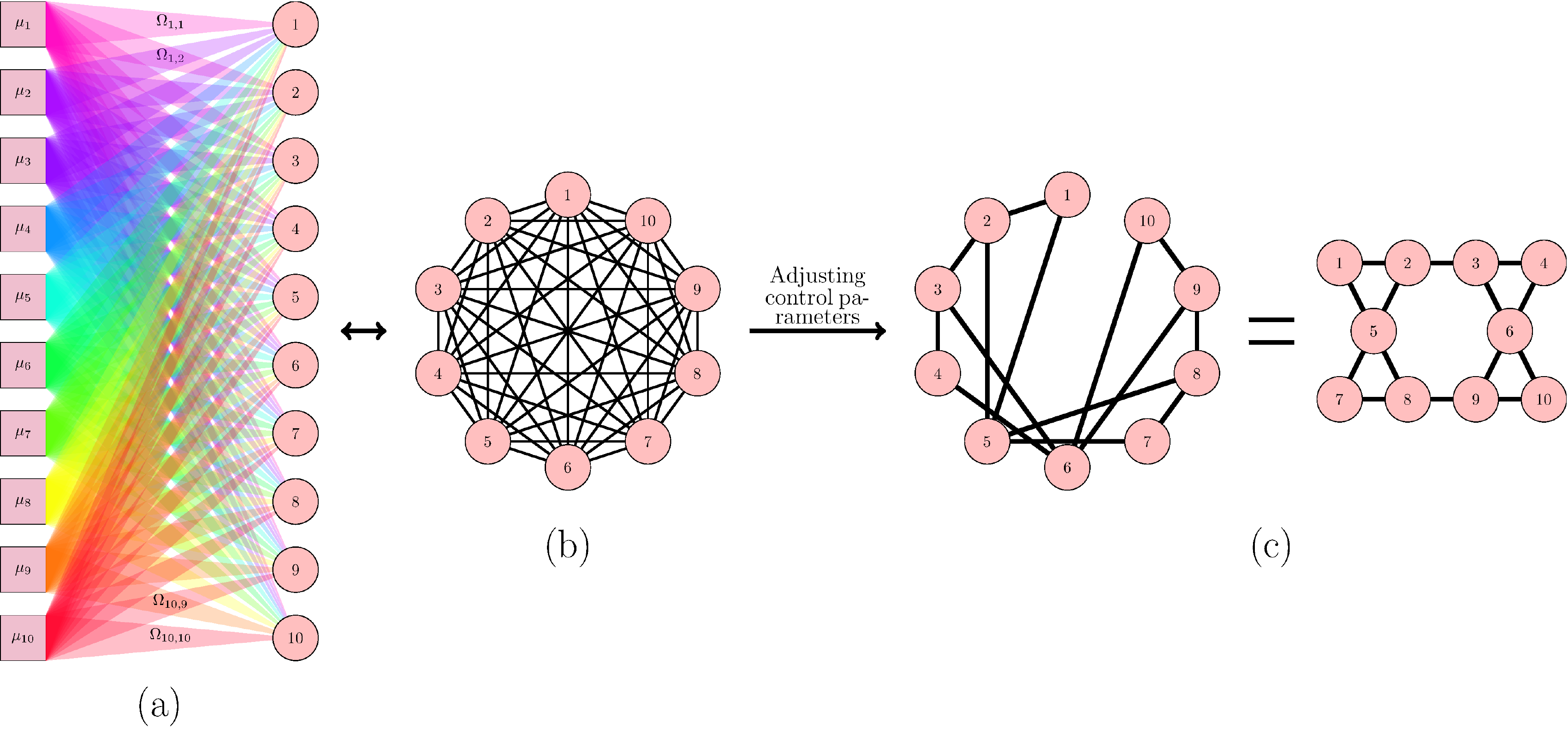}
    \caption{(a) Diagram of 10 linearly trapped ions individually addressed by 10 laser beams, of different frequencies, coming from the left. There is also a global beam on the trapped ions coming from the right, omitted from (a). The difference between the frequencies of the individually addressing beams and the global beam generate the beat-notes. These beat-notes have frequencies, $\mu_n$, also known as the Raman beat-note frequencies. To control the interaction between the ions, the Raman beat-note frequencies and Rabi frequencies, $\Omega_{i,n}$ are adjusted. In general, the resulting spin-phonon couplings generate pairwise interactions between all the ions forming a (b) fully connected lattice. By adjusting the control parameters, an \textit{arbitrary} lattice, e.g. (c) kagome lattice, can be generated.}
    \label{fig:Schematic}
\end{figure}

Trapped-ion spins can be made to interact with each other by virtually exchanging  coherent phonon excitations \cite{Cirac1995quantum,Sorensen1999quantum,Kim2009entanglement}. Spin-$\frac{1}{2}$ systems have been extensively investigated in recent experiments \cite{Zhang2017many-body,Zhang2017discrete}, and higher spin systems that can demonstrate novel phases of matter can also be implemented \cite{Senko2015realization}. Local spin operators ${\bf S}_i$ of different symmetries
and with different pairwise interactions, such as Ising, XY or XXZ, can effectively
be prepared to generate spin Hamiltonians, e.g., 
\begin{equation}
H = \sum_{i,j|i<j} J_{ij} \hspace{1mm} \big( {\bf S}_i^+ {\bf S}_j^-+ {\bf S}_i^- {\bf S}^{+}_j \big) . \label{eq:Hxy}
\end{equation}
Controlling the spin-phonon couplings enables the creation of the spin-spin interaction graph $J_{ij}$. For a system of $N$ ions, there are ${N(N-1)}/{2}$ possible unique pairwise interactions $J_{ij}$ in the Hamiltonian Eq.~(\ref{eq:Hxy}).
The spin-phonon couplings can be excited using radiation fields, and it is the control parameters of these radiation fields that must be determined in order to generate a target spin-spin interaction graph. 

Without the loss of generality, we assume a simulator made of $N$ trapped-ion hyperfine qubits or spin states. Two-photon coherent Raman transitions induced by counter-propagating laser beams are used to couple each of the $N$ ion-spins to $N$ collective vibrational or phonon states of the ion chain along a spatial direction. Hence, there are $N^2$ spin-phonon couplings that can be tuned to achieve a target $J_{ij}$ graph. This is illustrated in Fig.~(\ref{fig:Schematic}) for a chain of $N=10$ ions. Each ion is irradiated with $N$ bi-chromatic laser beams \cite{Sorensen1999quantum, Kim2009entanglement}. Each of these bi-chromatic beams interferes with a counter-propagating global laser beam (not shown in the figure), shining uniformly on all ions to produce  bi-chromatic `Raman beat-notes' at frequencies of $\omega_0\pm\mu_n$ ($n=1,2,\cdots,N$), $\omega_0$ being the qubit frequency. We will refer to $\mu_n$ as the beat-note frequency in this manuscript. As we will see in Eqs.(\ref{eq:Detuning_1}) and (\ref{eq:Detuning_2}) we will set each beat-note $\mu_n$ to be close to each of the $N$ collective phonon modes. The interaction between these $N$ laser beams with $N$ ions are quantified by $N^2$ `Rabi frequencies', $\Omega_{i,n}$ ($i=1,2,...,N$ denoting the ion index, and $n=1,2,...,N$ denoting the phonon mode index), which are related to the electric field of the laser beams at the ions, and the $N$ beat-note frequencies, $\mu_n$. These Rabi frequencies $\Omega_{i,n}$ are our control parameters. 
The spin-spin interaction graph produced by the spin-phonon couplings, $J_{ij}$, has the following dependence on the control parameters \cite{Korenblit2012quantum, Kim2009entanglement},
\begin{equation}
    J_{ij}(\Omega_{i,n},\mu_n) = 
    \sum_n^N\Omega_{i,n}\Omega_{j,n}\sum_m^N\frac{\eta_{i,m}\eta_{j,m}\omega_m}{\mu_n^2-\omega_m^2}.
    \label{eq:Jij}
\end{equation}
Here, $\eta_{i,m}=b_{i,m}\delta k\sqrt{{\hbar}/{2M\omega_m}}$ is the Lamb-Dicke parameter, $b_{i,m}$ is the phonon mode transformation matrix, $\delta k$ is the wave vector difference of the counter-propagating Raman lasers (which does not vary appreciably between different beat-note frequencies), $\omega_m$ are the phonon mode frequencies, and $M$ is the mass of a single ion. In order for the spin-models, such as Eq.~(\ref{eq:Hxy}), to be a good approximation of the full system Hamiltonian, direct phonon excitations must be avoided, which can be achieved by spacing the Raman beat-note frequencies $\mu_n$ from the normal mode frequencies \cite{Korenblit2012quantum}.
Without any loss of generality, we fix the Raman beat-note frequencies using
\begin{eqnarray}
    \mu_1 &=& \omega_1 + 0.1(\overline{\Delta\omega}),  \label{eq:Detuning_1} \\
    \mu_n &=& \omega_n + 0.1(\omega_{n-1}-\omega_n), \ \mathrm{for} \ n > 2,  \label{eq:Detuning_2}
\end{eqnarray}
where $\overline{\Delta\omega}$ is the mean separation between neighbouring phonon mode frequencies.

As can be deduced from inspection of Eq.~(\ref{eq:Jij}), the process of determining the control parameters given a target $J_{ij}$ is difficult due to the nonlinearity, in $\Omega_{i,n}$, of the equation.
Constrained nonlinear optimization schemes have been attempted in the past \cite{Korenblit2012quantum}.  However, we seek to explore more 
powerful methods that may be more efficient and scalable. To this end, we note that the the task of determining control parameters 
given a target $J_{ij}$ is analogous to an {\em inverse problem} -- commonly encountered in physics when one wishes to
estimate causal factors from a set of observations \cite{Engl1996regularization}.
In our case, the {\em forward problem} given by Eq.~(\ref{eq:Jij}), i.e.~determining 
the interaction matrix $J_{ij}$ given $\Omega_{i,n}$ and $\mu_n$, is generally straightforward and linear.  In contrast, the inverse problem
of calculating the set of $\Omega_{i,n}$ and $\mu_n$ that lead to a target $J_{ij}$ is highly non-linear, and in the worse case 
unstable, non-unique, or otherwise ill-posed.

Fortunately, there has been extensive research in the last several years using artificial neural networks and other machine learning algorithms
to attack such inverse problems
\cite{Adler2017solving,Arsenault2017projected,Fournier2018artificial}.  
The setting of such approaches are data driven - meaning that a large number of forward problems are first solved in order 
to produce a set of training data.  This data is used to train the parameters of a nonlinear function that can then be exposed to
data outside of the training set.  The hope is that this procedure will {\em generalize} well, meaning that a data set of 
reasonable (finite) size can be used to produce good results for $J_{ij}$ not contained in the training set.

\begin{figure}[t]
    \centering
    \includegraphics[width=\columnwidth]{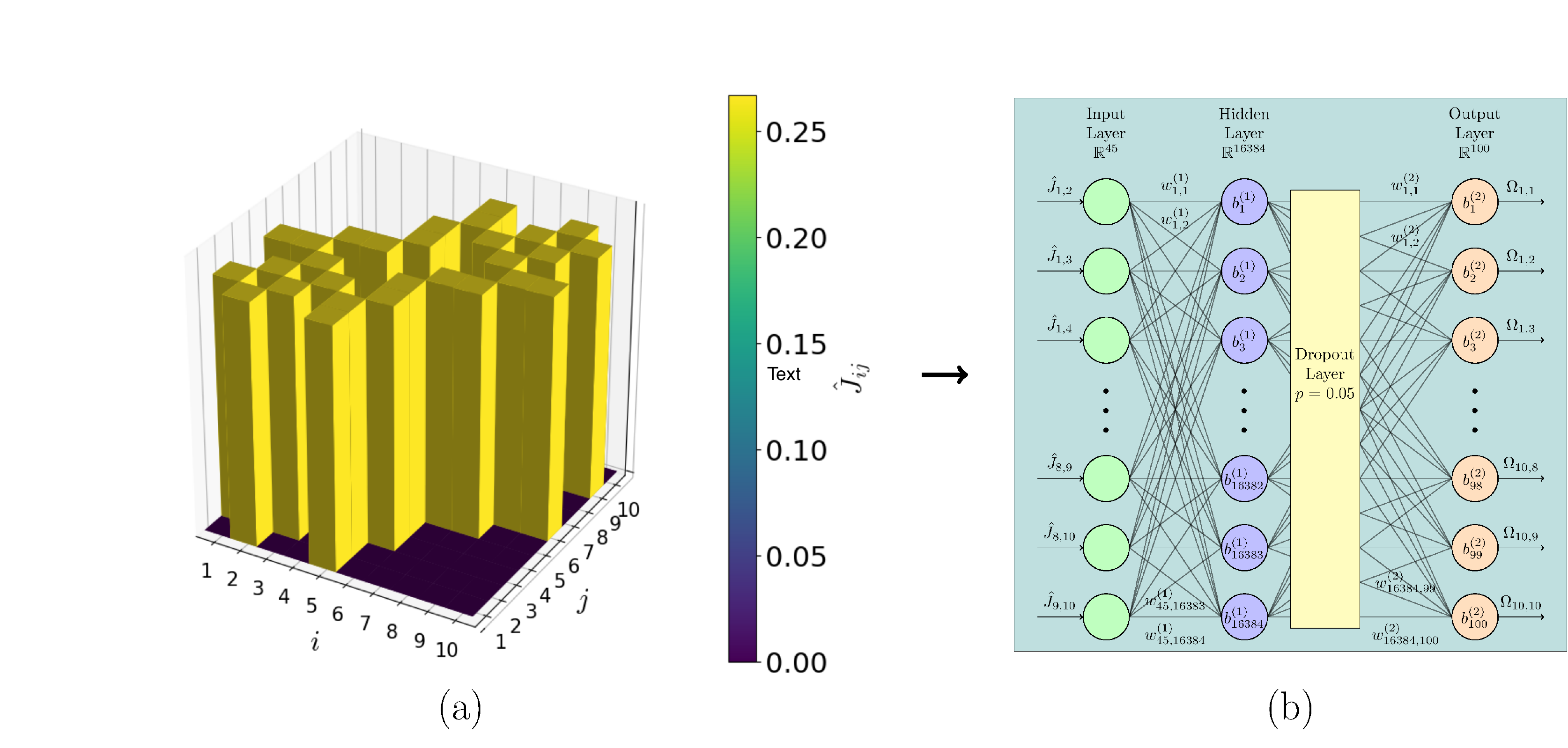}
    \caption{Diagram representative of the process to obtain the laser field control parameters for an \textit{arbitrary} interaction graph. Considering a system size of 10, the normalized interaction graph, e.g. (a) kagome lattice, $\hat{J}_{ij}$, is propagated through the proposed (b) neural network, as described in the text. After which, the neural network outputs the control parameters, i.e. Rabi frequencies, $\Omega_{i,n}$, required.}
    \label{fig:Neural net}
\end{figure}

In this paper we use a powerful artificial neural network scheme, adopted from previous applications to inverse problems in physics \cite{Fournier2018artificial} and slightly modified for our particular application - see Fig.~(\ref{fig:Neural net}).
The neural network takes as input a set of normalized unique (target) pairwise interactions, which we call $\hat{J}_{ij}$, defined as,
\begin{equation}
    \hat{J}_{ij} = \frac{J_{ij}}{||J||},
    \label{eq:NormJij}
\end{equation}
where $||J|| = \sum_{i,j|i<j}J_{ij}^2$, the typical vector norm on the unique pairwise interactions. 
Its outputs are the Rabi frequencies, $\Omega_{i,n}$.  
For the purposes of this study, we fix the Raman beat-note frequencies $\mu_n$ as {\em hyperparameters}, meaning that they
are adjusted outside of the data-driven training approach; in particular, they are adjusted according to Eqs.~(\ref{eq:Detuning_1}) and (\ref{eq:Detuning_2}).  
The internal structure of the neural network consists of an input layer, fully connected to a hidden layer with rectified linear unit (ReLU) activation, followed by a 5\% dropout layer and fully connected to an output layer with linear activation.
The size of the input of the network is dependent on the size of the $\hat{J}_{ij}$ matrix. As previously mentioned we only need to consider the entries above the diagonal, which gives an input size of ${N(N-1)}/{2}$. Empirically, we set the size of the hidden layer to be $16384$ as this value allowed the neural network to generalize well to unseen data for $N<50$. The network outputs the predicted values of $\Omega_{i,n}$, and so the size of the output is $N^2$. The linear activation for the output was chosen to allow for greater degrees of freedom for $\Omega_{i,n}$. These additional degrees of freedom comes from the ability to output negative values for $\Omega_{i,n}$, which corresponds to a phase difference between the individually addressing beam and the global beam.

As discussed, the forward problem of Eq.~(\ref{eq:Jij}) allows us to create an arbitrary amount of data with which to perform training.
In particular, we use random values of Rabi frequencies, drawn from a continuous uniform distribution from -1 to 1, to generate our training data.
During the training process, we optimize the neural networks parameters, using a method for stochastic optimization called the ADAM algorithm \cite{Kingma2014adam}, to minimize the following mean squared error (MSE) cost function,
\begin{equation}
    C(\Omega_{i,n},\hat{J}_{ij}) = \frac{2}{N(N-1)}\sum_{i,j|i<j}^N (\hat{\mathcal{J}}_{ij}(\Omega_{i,n}) - \hat{J}_{ij})^2.
    \label{eq:NNCost}
\end{equation}
Here, $\hat{\mathcal{J}_{ij}}$ is the normalized spin-spin interaction graph obtained by running the values of $\Omega_{i,n}$ output by
the neural network through the forward problem, Eq.~(\ref{eq:Jij}). Other hyperparameters must be set for the optimization; these include the training and validation dataset size, which are 45000 and 5000 respectively, an initial learning rate of $10^{-3}$, and a learning rate decay every 5 epochs of 90\%. We trained our neural network for a total of 100 epochs.

We implemented our machine learning algorithm with the Python language using Pytorch, an open source machine learning library \cite{paszke2017automatic}. 
For all of the results that we now discuss, 
the code was run on a Nvidia GTX 1060 graphics processing unit (GPU) with 6GB of video random access memory (VRAM).

\section{Results}
We consider a configuration of $N$ trapped ytterbium ions, $^{171}\mathrm{Yb}^{+}$,
addressed with Raman laser beam of mean wavelength 355 nm. The trapping potential is specifically set such that the equilibrium configuration of the trapped ions forms a linear chain and the normal mode frequencies (transverse to the ion chain) lie in the range of $2\pi\times$1 MHz to $2\pi\times$5 MHz. We train the neural network for the specified trapped ion chain. 
Here, we focus on the creation of interactions on regular lattices in one, two and three dimensions.  Such lattices are important for
quantum simulations of models in condensed matter, where physicists are typically interested in the behavior of
matter and materials at low energies, where crystalline structures form.
Of particular importance are so-called {\it frustrated} lattices, such as the two-dimensional triangular or kagome
lattices, which for certain interactions are not amenable to conventional (classical) computer simulations, e.g.~due to the infamous
quantum Monte Carlo ``sign problem''.  Outside of condensed matter, e.g.~in lattice gauge theories, interactions that encode regular lattice graphs are also important, where they often represent a discretization of space-time.  There, hypercubic geometries are used.  
In this section, we present the results of our neural network procedure for the creation of some typical lattices of 
interest to these and other areas of physics.

\subsection{Numerical results for $N=10$ ions}
\begin{figure}[t]
    \centering
    \includegraphics[width=0.9\columnwidth]{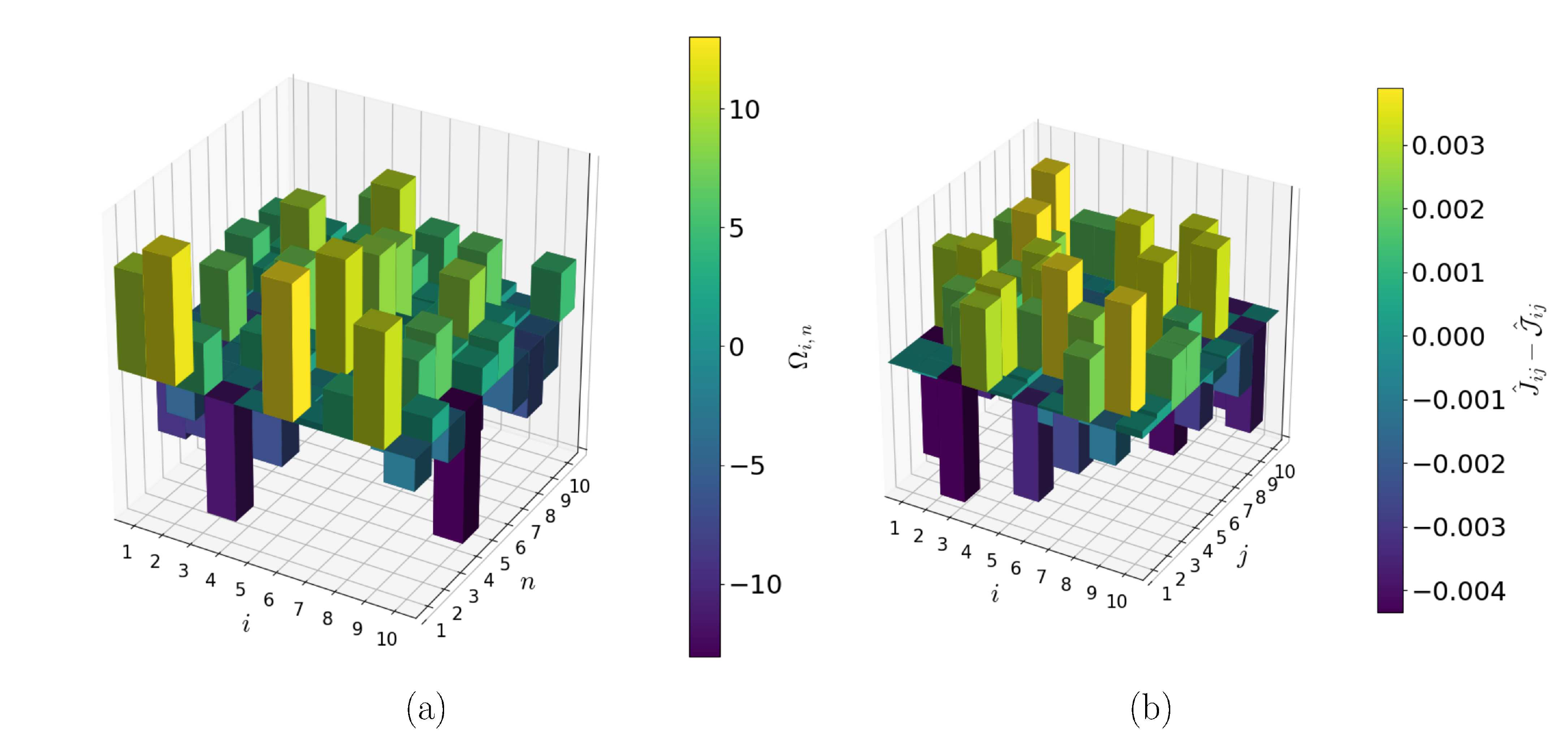}
    \caption{For the scenario proposed above, the neural network outputs the (a) Rabi frequencies, $\Omega_{i,n}$. Substituting $\Omega_{i,n}$ into Eq.~(\ref{eq:Jij}), we obtained the generated interaction graph, $\mathcal{J}_{ij}$. The generated normalized interaction graph, $\hat{\mathcal{J}}_{ij}$, matches the target normalized interaction graph, $\hat{J}_{ij}$, to a high precision, as can be deduced from the (b) difference between them. The similarity is further quantified by a similarity function $\mathcal{F}$, defined in Eq.~(\ref{eq:SimFn}), and shown in Table~\ref{tab:Similarity} for a range of target interaction graphs.}
    \label{fig:Rabi freqs}
\end{figure}

We first consider a linear trapped ion chain of $N=10$ ytterbium ions, and use our nerual network prodedure to 
generated interaction graphs, $\mathcal{J}_{ij}$, for several lattices.  
As an example, Fig.~(\ref{fig:Rabi freqs}) shows the results for the Rabi frequencies from the trained neural network when asked to produce a two-dimensional kagome lattice.
One can examine the difference between the generated and target interaction graph to gauge the quality of the reconstruction.  In the case of the kagome lattice on ten ions, the difference in the normalized interaction graphs lies in the third decimal place.

Once can explicitly quantify the trained neural network's ability to predict the control parameters for a variety of lattices of different geometries. 
For this, we define the following similarity function,
\begin{equation}
    \mathcal{F}(\hat{\mathcal{J}}_{ij},\hat{J}_{ij}) = \sum_{i,j|i<j}^N \hat{\mathcal{J}}_{ij} \ \hat{J}_{ij},
    \label{eq:SimFn}
\end{equation}
which quantifies the similarity between the normalized predicted and target spin-spin interaction graphs. The similarity function has a range of $[-1,1]$. When it has a value of 1, the (normalized) predicted and target spin-spin interaction graph are identical. When $\mathcal{F}=0$, the (normalized) predicted and target spin-spin interaction graph are ``orthogonal'', i.e. they are completely different. When $\mathcal{F}=-1$, the normalized predicted and target spin-spin interaction graph differ by a global negative sign. 

We examine the similarity function for a number of target graphs that are regular lattices of interest, e.g.~to condensed matter physics.
As shown in Table~\ref{tab:Similarity}, the similarity of the simulated spin-spin interaction graph achieved from the neural network and the target spin-spin interaction graph, for a variety of typical lattice geometries, is well above 0.999. 
As we discuss in Section \ref{ExpSec}, we believe that this is sufficiently accurate for real experimental ion traps, 
where other sources of error such as crosstalk are expected to provide larger sources of error.

\begin{table}[t]
    \begin{tabular}{cc|c}
        Lattice & &$\mathcal{F}$ \\
        \hline \\
        (a) & \includegraphics[height=0.12in]{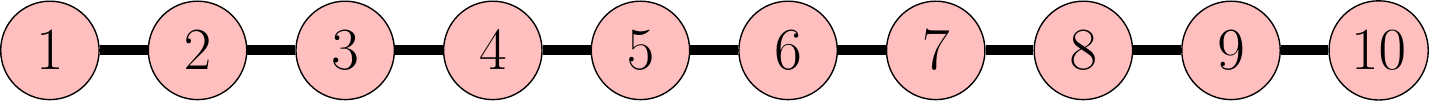} & 0.99988 \\ 
        (b) & \includegraphics[height=0.32in]{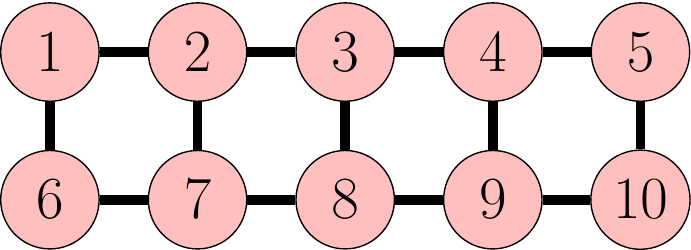} & 0.99983 \\ 
        (c) & \includegraphics[height=0.32in]{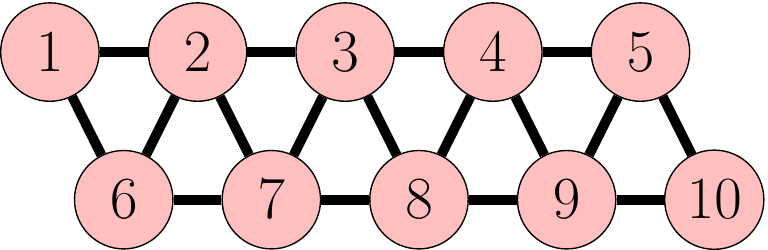}  & 0.99989 \\ 
        (d) & \includegraphics[height=0.56in]{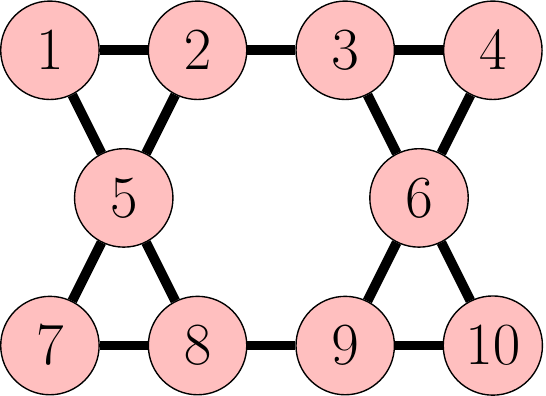}  & 0.99989 \\ 
        (e) & \includegraphics[height=0.64in]{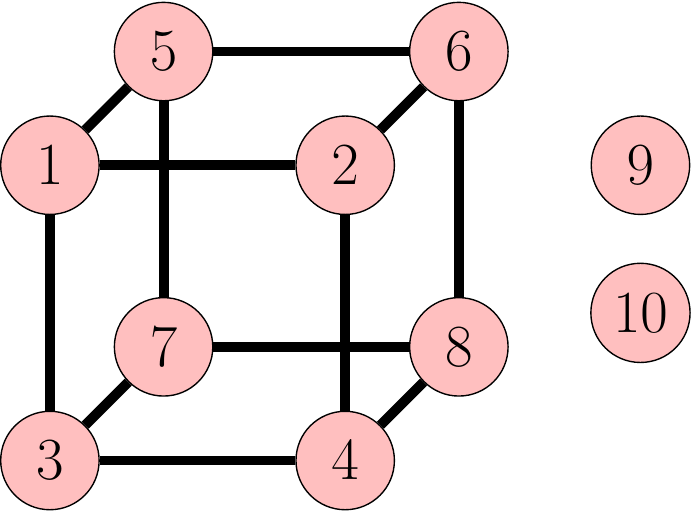}  & 0.99983 \\ 
        (f) & \includegraphics[height=0.32in]{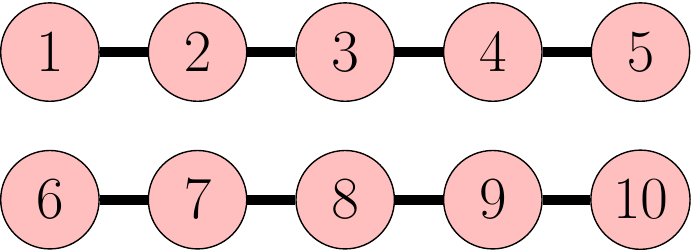} & 0.99988 \\
    \end{tabular}
    \caption{Similarity between the neural network produced and target normalized interaction graph, $\mathcal{F}$, for various lattice geometries, (a) linear chain, (b) square lattice, (c) triangular lattice, (d) kagome lattice, (e) cubic lattice and (f) 2 separate linear chains. Notice that $\mathcal{F} > 0.999$ for the above lattices. As a typical crosstalk of $\epsilon=0.01$, defined by Eqs.~(\ref{eq:XTalk_1}), (\ref{eq:XTalk_2}) and (\ref{eq:XTalk_3}), has a similarity of , $0.998 < \mathcal{F} < 0.999$, with the generated interaction graph, the neural network is sufficiently accurate, i.e. it is not the limiting factor in regards to accuracy.}
    \label{tab:Similarity}
\end{table}

\subsection{Numerical results for $N<50$ ions}
In the last section, we have shown that our proposed method works well for $N=10$ ions.  We now explore the scaling of the method with larger number of ions more systematically.  We first adjust the trapping strengths in the $x$ and $z$ directions, $\omega_x, \ \omega_z$, appropriately, i.e. such that the ions satisfy the aforementioned condition on its equilibrium configuration and the range of the transverse mode frequencies. 

\begin{figure}[t]
    \centering
    \includegraphics[width=\columnwidth]{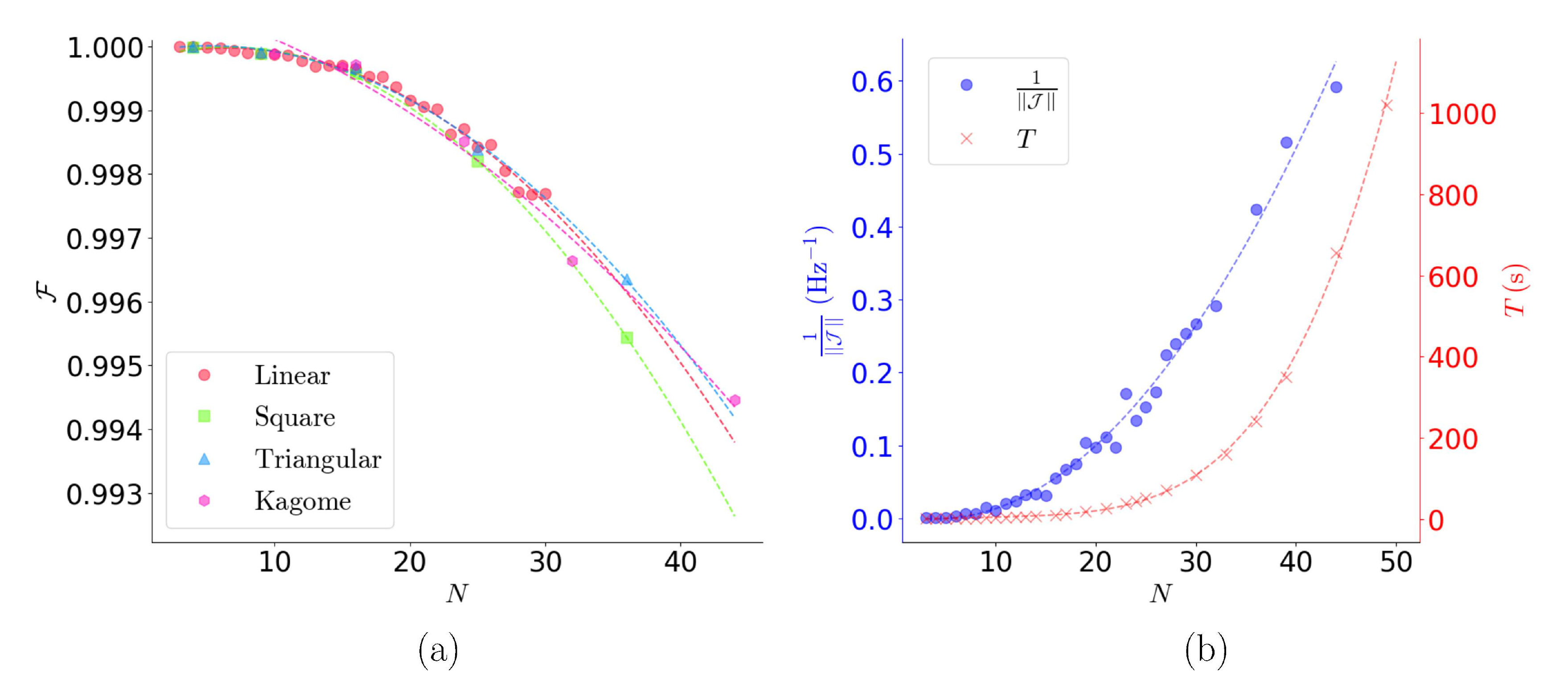}
    \caption{Investigation into the scaling of the proposed method with the number of ions, $N$, where dotted lines represent polynomial fits. The (a) similarity, between the neural network produced and target normalized interaction graph, $\mathcal{F}$, for various lattice geometries: linear chain, square lattice, triangular lattice and kagome lattice, decays with $\mathcal{O}(N^{2})$. The (b) time required to train an epoch in seconds, $T$, grows with $\mathcal{O}(N^4)$ and the (b) interaction strength, $||\mathcal{J}||$, of a linear chain with a constraint on the total power of, $\sum_{i,n}|\Omega_{i,n}| / 2 \pi = 1 \mathrm{MHz}$, scales inversely with $\mathcal{O}(N^{2})$.}
    \label{fig:Scaling}
\end{figure}

In Fig.~(\ref{fig:Scaling}), we begin by investigating the scaling of the similarity function of a variety of lattice geometries.  For $N<50$, we observe that $\mathcal{F}$ remains above 0.99 and the decay is a good fit to a function quadratic in the number of ions, $\mathcal{O}(N^2)$. Next, we characterize the scaling of the training time of a single epoch. We note a marked increase in the training time for $N$ approaching 50 ions, which has a good fit to a fourth-order polynomial $\mathcal{O}(N^4)$. This increase in training time is primarily due to the VRAM requirements for the parallel calculation of spin-spin interaction graphs in accordance to Eq.~(\ref{eq:Jij}).  With relatively modest increases in hardware, we believe it would be straightforward to push such calculations for ion numbers significantly larger than $N=50$.

\subsection{Experimental considerations} \label{ExpSec}

As established above, our neural network was able to determine the control parameters required to simulate the geometries of arbitrary spin models. However, the strength of the interaction in such spin models have not been accounted for. Theoretically, as shown in Eq.~(\ref{eq:Jij}), by increasing/decreasing the magnitude of the Rabi frequencies, we are able to manipulate the strength of the interaction in the spin models arbitrarily. In practice, the laser beams used to drive the spin-phonon couplings have finite power, and this leads to a constraint on the magnitude of the Rabi frequencies. Consequently, this would limit the maximum strength of the interactions, dependant on the size of the system. In Fig.~(\ref{fig:Scaling}), we investigate the scaling of this limit on interaction strength and subsequently determined that the inverse interaction strength is a good fit to a quadratic function in the number of ions, $\mathcal{O}(N^2)$, in the presence of a total power constraint on the laser beams.

Another relevant constraint arises from the assumptions required for the validity of Eq.~(\ref{eq:Jij}). As the spin-spin interactions are mediated by spin-phonon interactions, phonon states naturally participate in the interaction and, in general, cannot be ignored. Consequently, the system is not adequately described by the pure spin-spin Hamiltonian. However, if the Raman beat-note frequencies, $\mu_n$, and Rabi frequencies, $\Omega_{i,n}$, satisfy the following constraints, $|\mu_n - \omega_m| >> \eta_{i,m}\Omega_{i,n}$, the phonon states are only {\it virtually excited} and can be {\it adiabatically eliminated} \cite{Korenblit2012quantum,Kim2009entanglement}. In our approach, these constraints have not been strictly enforced, unlike in previous attempts at the problem \cite{Korenblit2012quantum}, as such, the validity of the solutions, produced by the neural network, requires explicit verification. Considering the total Rabi frequency limit as specified in Fig.~(\ref{fig:Scaling}), we found that the aforementioned solutions satisfy the constraints in question and excite approximately 0.0005 phonons. Therefore, they are eligible solutions for experiments.

As can be deduced from Eq. (\ref{eq:Jij}), the previously mentioned constraints interplay to create an upper bound on the interaction strengths that are simulatable on the system.  As such, obtaining larger interaction strengths require the relaxation of one of the constraints. In practice, there are situations where obtaining a higher powered laser is not an option. In such a scenario, the frequencies, $\mu_n$, could be set closer to the phonon mode frequencies, $\omega_m$, to obtain larger interaction strengths at the expense of more phonon excitations.

Finally, we consider experimental errors that could occur when individually addressing the ions using the laser field.  For example, an important error is crosstalk between the laser beams when individually addressing the ion chain with multiple frequencies. To investigate the effect of crosstalk on the resulting interaction graph, $\mathcal{J}_{ij}$, we define a crosstalk magnitude of $\epsilon$ as,
\begin{eqnarray}
    \Omega_{i,n}^{\mathrm{crosstalk}} &=& \Omega_{i,n} + \epsilon \ (\Omega_{i-1,n} + \Omega_{i+1,n}), \ \mathrm{for} \ 1 < i < 10, \label{eq:XTalk_1} \\
    \Omega_{1,n}^{\mathrm{crosstalk}} &=& \Omega_{1,n} + \epsilon \ \Omega_{2,n}, \label{eq:XTalk_2} \\
    \Omega_{10,n}^{\mathrm{crosstalk}} &=& \Omega_{10,n} + \epsilon \ \Omega_{9,n}.
    \label{eq:XTalk_3}
\end{eqnarray}
We use these to define an error function,
\begin{equation}
    \mathcal{E} = \frac{||\mathcal{J}(\Omega_{i,n}^{\mathrm{crosstalk}}) - \mathcal{J}(\Omega_{i,n})||}{||\mathcal{J}(\Omega_{i,n})||}.
    \label{eq:ErrFn}
\end{equation}
After computing this quantity, our numerical results show that the error in the interaction graph, $\mathcal{E}$, scales linearly with the crosstalk magnitude, $\epsilon$. In addition, the crosstalk also affects the distribution of the interactions, i.e.~the normalized interaction graph, $\hat{\mathcal{J}}_{ij}$. Considering a typical crosstalk of 1\%, corresponding to $\epsilon=0.01$, we note that the change in the distribution due to the crosstalk is larger than the difference in the distribution of the neural network generated and target interaction graph, i.e. $\mathcal{F}(\hat{\mathcal{J}}^{\mathrm{crosstalk}}_{ij},\hat{\mathcal{J}}_{ij}) < \mathcal{F}(\hat{\mathcal{J}}_{ij},\hat{J}_{ij})$. Therefore, the crosstalk between the laser beams would be the dominant contributor to the error in the distribution of the interaction graph.  Our neural network results are thus sufficiently accurate for the conditions expected in a realistic experimental situation.

\section{Conclusion and outlook}

We have demonstrated how machine learning can be utilised for programming a trapped ion quantum simulator for the realization of spin models with arbitrary interaction graphs, $J_{ij}$.
Our approach employs a feed forward neural network to solve the inverse problem associated with
determining laser control parameters, given as input an interaction graph of interest.  Solutions obtained in
this manner are easy to compare to the target $J_{ij}$, since the associated forward problem is
trivial, allowing for precise similarity comparisons for any  example of interest.

Contrary to previous approaches  \cite{Korenblit2012quantum}, our machine learning technique searches for an approximate relationship between the spin model's spin-spin interaction graph and the $N^2$ control parameters, which allows for an approximate solution to the inverse problem.  We observe that solutions found this way are ``stable''.  That is, small perturbations to the solutions -- of the order of magnitude expected experimentally -- do not drastically affect the results.  In addition, for graphs representing typical lattices in one, two and three dimensions, we find a favorable scaling of the similarity comparison up to 50 qubits.  
Since our machine learning procedure is implemented on only a single GPU workstation, this 
suggests that with moderate additional resources, lattices for hundreds of spins should easily be within reach.

As simulators and other near-term quantum devices continue to grow towards hundreds of qubits, a number of difficult experimental problems in design, control, and verification will require practical solutions to achieve scaling.
With the advent of a suite of machine learning techniques as powerful numerical tools, with applications across a spectrum of difficult problems of relevance to experimental quantum hardware,
we expect technologies like neural networks to become an integral part of every level of the stack of quantum simulators in the future.

\section*{Acknowledgements}
We thank M.~Beach, A.~Golubeva, B.~Kulchytskyy, R.~Luo and G.~Torlai for important discussions.  
The authors would like to thank TQT (CFREF) for financial support.
RGM is also supported by NSERC, the Canada Research Chair program, and the Perimeter Institute for Theoretical Physics. RI is also supported by Innovation, Science and Economic Development Canada (ISED), NSERC, Province of Ontario, IQC and University of Waterloo. Research at Perimeter Institute is supported in part by the Government of Canada through ISED and by the Province of Ontario through the Ministry of Economic Development, Job Creation and Trade.

\bibliography{ref}

\end{document}